\pdfoutput=1
\documentclass[10pt,journal]{IEEEtran}

\usepackage{cite}

\usepackage[pdftex]{graphicx}
\graphicspath{{images/}}

\usepackage[cmex10]{amsmath}
\usepackage{bm}
\usepackage{array}
\usepackage[caption=false,font=normalsize,labelfont=sf,textfont=sf]{subfig}
\usepackage{url}

\usepackage[T1]{fontenc}
\usepackage[utf8]{inputenc}
\usepackage[english]{babel}

\begin{document}

\title{A Topologically-informed Hyperstreamline\\Seeding Method for Alignment Tensor Fields}

\author{Fred~Fu and~Nasser~Mohieddin Abukhdeir
\IEEEcompsocitemizethanks{\IEEEcompsocthanksitem F. Fu and N.M. Abukhdeir are with the Department of Chemical Engineering, University of Waterloo, Waterloo, Ontario, Canada.\protect\\
E-mail: ffu@uwaterloo.ca, nmabukhdeir@uwaterloo.ca}
\thanks{}}

\markboth{IEEE~TRANSACTIONS~ON~VISUALIZATION~AND~COMPUTER~GRAPHICS, VOL. XX, NO. X, MONTH/YEAR}%
{Shell \MakeLowercase{\textit{et al.}}: Bare Demo of IEEEtran.cls for Computer Society Journals}

\IEEEcompsoctitleabstractindextext{%
\begin{abstract}
A topologically-informed method is presented for seeding of hyperstreamlines for visualization of alignment tensor fields.
The method is inspired by and applied to visualization of nematic liquid crystal (LC) reorientation dynamics simulations.
The method distributes hyperstreamlines along domain boundaries and edges of a nearest-neighbor graph whose vertices are degenerate regions of the alignment tensor field, which correspond to orientational defects in a nematic LC domain.
This is accomplished without iteration while conforming to a user-specified spacing between hyperstreamlines and avoids possible failure modes associated with hyperstreamline integration in the vicinity of degeneracies of alignment (orientational defects).
It is shown that the presented seeding method enables automated hyperstreamline-based visualization of a broad range of alignment tensor fields which enhances the ability of researchers to interpret these fields and provides an alternative to using glyph-based techniques.
\end{abstract}

}

\maketitle

\IEEEdisplaynotcompsoctitleabstractindextext
\IEEEpeerreviewmaketitle

\section{Introduction}

\IEEEPARstart{S}{imulation-based} research of liquid crystalline (LC) phases has played a key role both in the contribution to our fundamental understanding of these phases and to engineering of LC devices.
Liquid crystal phases, or mesophases, behave like disordered liquids at high temperatures, but upon cooling, transition to a lower symmetry liquid-like phase which possesses some degree of phase order.
The most simple of the LC phases is the nematic phase, which possesses some degree of orientational order at the molecular scale.
This orientational order is theoretically characterized using a second order symmetric traceless tensor $\bm{Q}$, the alignment tensor \cite{Sonnet1995}.
Applications of nematic LCs are pervasive in our daily lives ranging from LC-based displays (LCDs) to biological systems \cite{Rey2010}.

Resolution of the nanoscale structure and dynamics of nematic domains is challenging for experimental analysis, and thus simulation-based approaches are frequently employed both in fundamental and applied science.
The theoretical bases of these simulations have progressed from simple, but visually intuitive, vector field-based approximations of LC orientational order to more descriptive alignment tensor theory \cite{deGennes1995,Sonnet1995}.
Alignment tensor theory is more descriptive in that it captures degeneracies in alignment, orientational defects, and phase transition.
One of the persistent challenges resulting from using alignment tensor theory is that resulting three-dimensional transient simulation data have proven difficult to interpret.

Approaches to visualization of nematic alignment tensor fields have, until recently, resorted to simplifications such as extracting the major eigenvector of the tensor and visualizing it as a vector field \cite{Sparavigna1999,Zhu2002,Copar2013}.
Vector field visualization methods, especially streamline methods, are well-studied in the literature, and several methods for creating high quality streamline placements for two-dimensional data have been proposed \cite{Turk1996,Jobard1997,Verma2000,Mebarki2005,ChenY2007,ChenG2007,Wu2010}.
These include an image-guided algorithm using low-pass filtering \cite{Turk1996}, an approach based on separation distances \cite{Jobard1997}, a method that using templates for different types of critical points \cite{Verma2000}, a farthest point seeding strategy \cite{Mebarki2005}, similarity-guided streamline placement \cite{ChenY2007}, and topology-aware streamline placement \cite{Wu2010}. 
A thorough review of these streamline seeding strategies can be found in \cite{McLoughlin2010}.
However, while vector field approximations of alignment tensor fields enable visualization using standard streamline approaches, there are two significant drawbacks. First, degeneracies in alignment that are frequently present in alignment tensor fields result in singularities in vector fields. Second, much information is lost through the vector field approximation including the degree (or magnitude) of alignment and the presence of multiple alignment axes (biaxiality).

\begin{figure}[!t]
\centering
\subfloat[]{\includegraphics[width=0.45\linewidth]{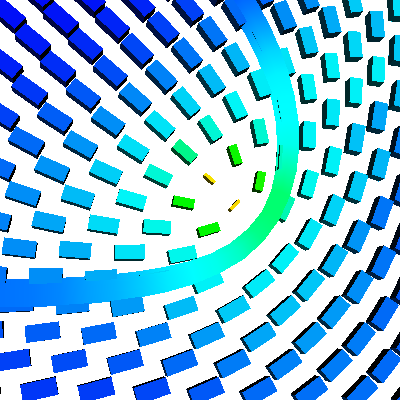}}
\hfil
\subfloat[]{\includegraphics[width=0.45\linewidth]{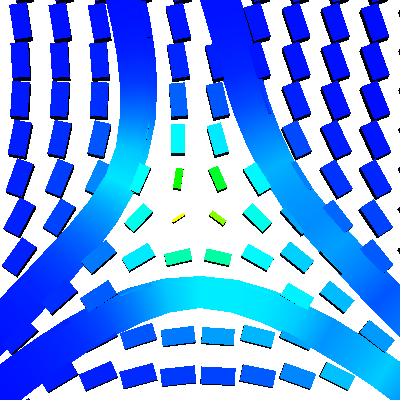}}
\caption{Two types of orientational defects commonly observed in nematic liquid crystals visualized with rectangular glyphs and hyperstreamlines: (a) a $+$ orientational defect and (b) a $-$ orientational defect.}
\label{fig:overview}
\end{figure}

As a result, recent advances have been made using tensor glyph methods (Figure \ref{fig:overview}), particularly the work by Jankun-Kelly and Mehta \cite{Jankun-Kelly2006,Slavin2006} which improves upon standard tensor glyph visualization by applying superellipsoids rather than using conventional glyph shapes. In general, improvements in tensor field visualization have resulted from the desire to interpret diffusion tensor imaging data for MRI \cite{Vilanova2006}.
More recently, Callan-Jones et al. \cite{Callan-Jones2006} employed streamsurfaces and streamtubes \cite{Zhang2003} to nematic domains that include topological defects in orientation \cite{Kleman1982} (shown in Figure \ref{fig:overview}), or disclinations, using Westin metrics to characterize the alignment tensor field, although this method relies on culling of computed streamlines in order to refine the visualization.
An adaptive streamtube seeding algorithm incorporating tensor dissimilarity measures also exists \cite{Weldeselassie2007}. 
Asymmetric tensor field visualization has also been studied \cite{Zhang2009, Chen2011}, although as the alignment tensor is symmetric, these methods cannot be directly adapted.

\begin{figure}[!]
\centering
\includegraphics[width=0.9\linewidth]{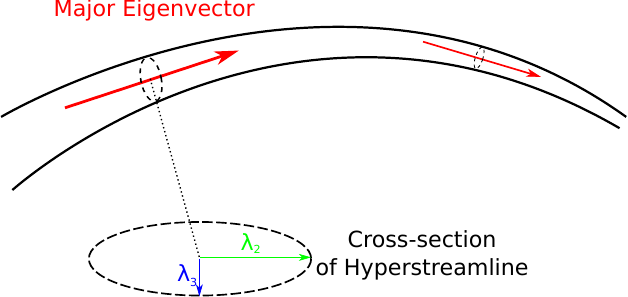}
\caption{Schematic of a hyperstreamline demonstrating its direction and cross-section.}
\label{fig:hyperstreamline}
\end{figure}

Hyperstreamline visualization of tensor fields \cite{Delmarcelle1993} (Figure \ref{fig:hyperstreamline}) is an alternative approach to glyph-based techniques.
Analogous to streamline visualization of vector fields, hyperstreamlines are enhanced such that, in addition to direction, they have volume.
This enables simultaneous visualization of all eigenvalues and eigenvectors of an alignment tensor field.
They are constructed by first computing a streamline using the major eigenvector field:
\begin{equation}\label{eqn:streamline}
	\frac{\mathrm{d}\bm{r}}{\mathrm{d}s} = \bm{n}(\bm{r})
\end{equation}
where $\bm{r}(s )$ is the position of the streamline, $s$ is the arc length along the streamline, and $\bm{n}$ is the major eigenvector of the alignment tensor field.
Using this streamline as a template, a hyperstreamline is then formed by rendering an elliptic cylinder such that the major/minor axis is aligned with the secondary/tertiary eigenvector field (of the alignment tensor).
The lengths of the major/minor elliptic axes are specified by the magnitude of the secondary/tertiary eigenvalue, shown in Figure \ref{fig:hyperstreamline}.
Thus, in contrast to streamlines, hyperstreamlines incorporate all information quantified by the alignment tensor in a higher-dimensional form than that of tensor glyphs.

As with streamline visualizations \cite{Verma2000}, one of the major challenges of employing hyperstreamlines is that existing seeding methods are either (i) simplistic (uniform spatial distributions) which result in difficult to interpret visualizations \cite{Zheng2004} or (ii) are complex iterative algorithms which are impractical for large three-dimensional transient datasets.
Additionally, alignment tensor fields frequently include degeneracies in their major eigenvector fields such that $\bm{n}(\bm{r})$ in eqn \ref{eqn:streamline} is not well-defined.
In this context, the three main objectives of this work are to develop a hyperstreamline seeding method such that:
\begin{enumerate}
    \item generation of seed points results in an \textit{approximately} well-distributed hyperstreamline visualizations.
    \item generated seed points avoid computation of hyperstreamlines which intersect areas of degenerate alignment.
    \item iteration is not required so that the method is feasible for use in the visualization of large three-dimensional transient datasets.
    \item a priori knowledge of the alignment tensor field, specifically the type of orientational degeneracies that are present, is not required.
\end{enumerate} 
Methods do exist to identify and avoid tensor degeneracies \cite{Zheng2004}, but they are computationally complex and preclude the use of functionality in existing visualization libraries, specifically the Visualization Toolkit (VTK) \cite{VTK}.
Recent work has shown that utilization of the orientational topology of alignment tensor fields could result in significant gains \cite{Verma2000,Tricoche2001,Ye2005}.

In this work, a seeding method is presented for the visualization of alignment tensor fields using hyperstreamlines in a way that incorporates topological information.
Degeneracies in alignment and orientational defects are used to form a spatial graph with edges determined from nearest-neighbor triangulation.
The vertices and edges are then used as a template for seeding in a way that, without resorting to iteration/pruning, both approximates an optimal distribution of hyperstreamlines throughout the domain and avoids hyperstreamline computation in the vicinity of defects.
The method is evaluated on a representative set of two-dimensional alignment tensor fields resulting from continuum simulations of nematic LC reorientation dynamics.

The paper is organized as follows: the alignment tensor and simulation method are described in Section \ref{sec:background}, the topologically-informed seeding method is presented in Section \ref{sec:methods}, results of applying the method to various two-dimensional alignment tensor fields are presented and discussed in Section \ref{sec:res_and_disc}, and conclusions are made in Section \ref{sec:conclusions}.

\section{Background}\label{sec:background}

\subsection{The Alignment Tensor}\label{sec:alignment}

The alignment tensor $\bm{Q}$ is a real second-order symmetric-traceless tensor and thus has distinct eigenvectors and real eigenvalues.
A symmetric tensor can be decomposed using its eigenvectors ($\bm{n}$, $\bm{m}$, $\bm{l}$) and eigenvalues ($\lambda_n$, $\lambda_m$, $\lambda_l$) using Dyadic/Gibbs tensor notation \cite{Cajori1993}:
\begin{equation}
\bm{Q} = \lambda_n \bm{nn} + \lambda_m \bm{mm} + \lambda_l \bm{ll}
\end{equation}
As in ref. \cite{Callan-Jones2006} we introduce the modified alignment tensor $\bm{D}$ with non-negative eigenvalues to simplify implementation of the method,
\begin{equation}
\bm{D} = \bm{Q} + \frac{1}{3}\bm{\delta}
\end{equation}
where $\lambda'_n \ge \lambda'_m \ge \lambda'_l$ are the eigenvalues of $\bm{D}$, the eigenvectors of $\bm{D}$ remain the same as $\bm{Q}$, and $\bm{\delta}$ is the identity tensor.

A useful decomposition of the modified alignment tensor $\bm{D}$ uses Westin metrics \cite{Westin1997}: the isotropy measure $c_s$, the linear anisotropy measure $c_l$, and the planar anisotropy measure $c_p$,
\begin{equation}
c_s,c_l,c_p\in [0,1], c_s + c_l + c_p = 1 
\end{equation}
The alignment tensor characterizes three general types of alignment:
\begin{enumerate}
\item $c_s \approx 1$ -- isotropy or no preferred alignment.
\item $c_l \approx 1$ -- uniaxial alignment along $\bm{n}$.
\item $c_p \approx 1$ -- biaxial alignment along $\bm{n}$ and $\bm{m}$.
\end{enumerate}

\begin{figure*}[!]
\centering
\includegraphics[width=0.9\linewidth]{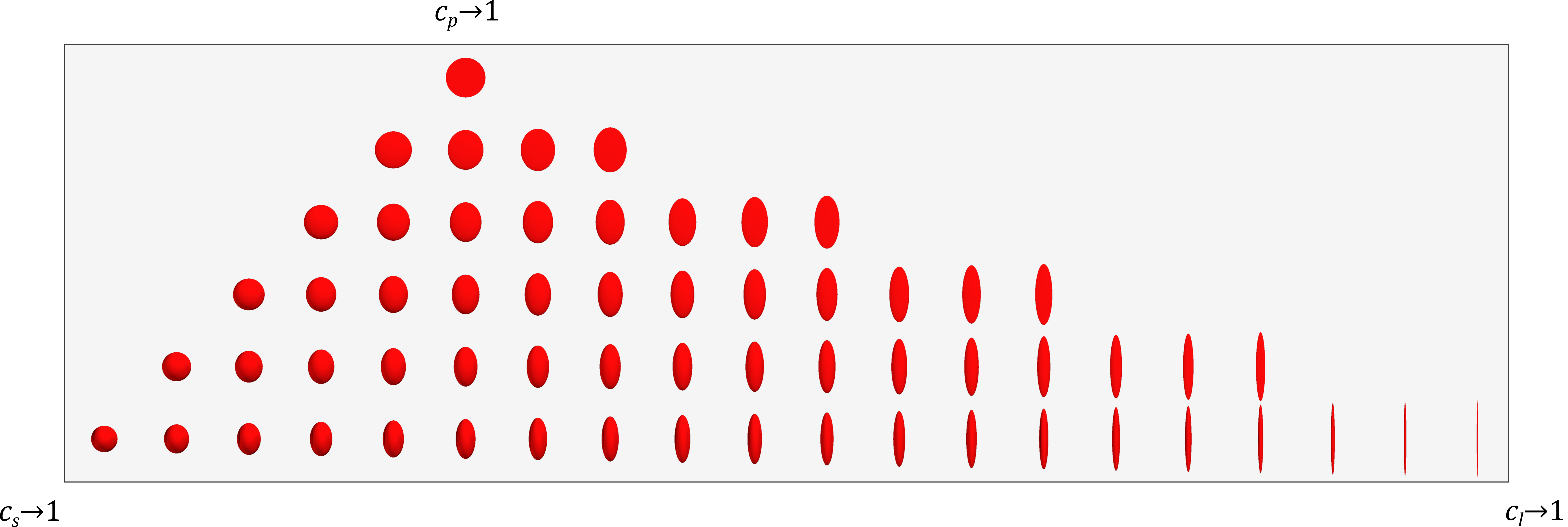}
\caption{Tensor ellipsoid representations of the modified alignment tensor $\bm{D}$ for differing Westin metrics values  corresponding to: (a) no alignment (isotropic) ($c_s \approx 1$), (b) uniaxial alignment ($c_l \approx 1$), and (c) biaxial alignment $c_p \approx 1$.}
\label{fig:westin}
\end{figure*}

The relationship between the eigenvalues of $\bm{D}$ and Westin metrics of are \cite{Callan-Jones2006}:
\begin{equation}
c_l = \lambda'_1 - \lambda'_2, c_p = 2 (\lambda'_2 - \lambda'_3), c_s = 3 \lambda'_3
\end{equation}
Figure \ref{fig:westin} shows schematic examples of the three types of alignment in terms of both Westin metrics.

\subsection{Nematic Reorientation Dynamics}\label{sec:nematic}

Alignment tensor fields analyzed in this work are generated through simulations of nematic reorientation dynamics in the absence of flow.
Nematic dynamics equations are described in detail in ref. \cite{Rey2010}, and are summarized here.
A gradient flow model is used to simulate dynamics of the alignment tensor \cite{Desai2009,Bhattacharjee2010},
\begin{equation}\label{eqn:langevin}
\frac{\partial \bm{Q}}{\partial \mathrm{t}} = -\bm{\Gamma} : \frac{\delta F}{\delta \bm{Q}}
\end{equation}
where $F$ is the total free energy of the domain and the kinetic coefficient $\bm{\Gamma}$ is defined to preserve the symmetry and traceless properties of the alignment tensor.

The free energy density of the nematic domain used is given by the Landau-de Gennes model \cite{deGennes1995,Barbero2000},
\begin{eqnarray}\label{eqn:free_en_dens}
f &=&\frac{1}{2} a \left(\mathbf{Q} : \mathbf{Q}\right) - \frac{1}{3} b \left(\mathbf{Q}\cdot\mathbf{Q}\right) : \mathbf{Q} + \frac{1}{4} c \left(\mathbf{Q} : \mathbf{Q}\right)^2 \nonumber\\
&& + \frac{1}{2} L_1 (\mathbf{\nabla} \mathbf{Q} \vdots \mathbf{\nabla} \mathbf{Q} ) 
\end{eqnarray}
where material constants $a/b/c$ characterize the stability of the aligned LC (nematic) phase and $L_1$ characterizes its orientational elasticity.
Integration of the free energy density over the domain volume ($V$) results in the total free energy $F$,
\begin{equation}
    F = \int_{V} f\mathrm{d}V
\end{equation}

\section{Methods}\label{sec:methods}

The presented method is described for two-dimensional alignment tensor fields. The method is composed of three steps:
\begin{enumerate}
\item Identification of a topological template of the field from the domain boundary and orientational defects (if present).
\item Computation of an approximation of the optimal distribution of seed points guided by the topological template. 
\item Computation of hyperstreamlines at every seed point using the topological template while avoiding regions with orientational defects.
\end{enumerate}
The method requires only one parameter from the user, $l_s$, the desired spacing between hyperstreamlines in the final visualization.
For alignment tensor fields corresponding to nematic LC domains, there also exists a physical length scale over which the alignment tensor can vary, $l_n = \sqrt{\frac{L_1}{a}}$ \cite{deGennes1995}, which is used as a basis for choosing $l_{s}$ in the present work.

\subsection{Identification of a Topological Template}\label{sec:topo_template}

The alignment tensor field is first analyzed for the presence of degeneracies/defects in alignment (see Figure \ref{fig:overview}) through identifying regions with biaxial alignment ($c_p \approx 1$), as described in Section \ref{sec:alignment}).
From this analysis, a set of points in space $\mathcal{D}$ is generated which provides topological information about the domain of the domain.

Once $\mathcal{D}$ is determined, an undirected graph $\mathcal{G}$ is formed whose vertices are composed of $\mathcal{D}$ and whose edges relate each point to its nearest neighbor in space (via Delaunay triangulation).
The resulting graph is referred to as the \emph{topological template}. 
Using this template, curves $\bm{r}_{i}(s)$ (where $i$ is used arbitrarily to index each curve) are defined from a combination of the line segments/graph edges connecting nearest-neighbor defects and circles enclosing each defect/graph vertex.
In the case of well-aligned domains, the set $\mathcal{D}$ could be empty.
In this case, the orientational topology of the domain is completely described by the alignment at the domain boundary, which is used to define the curves $\bm{r}_{i}(s)$.

\subsection{Seed Distribution}\label{sec:seed_distribution}

\begin{figure}[!t]
\centering
\subfloat[]{\includegraphics[width=0.45\linewidth]{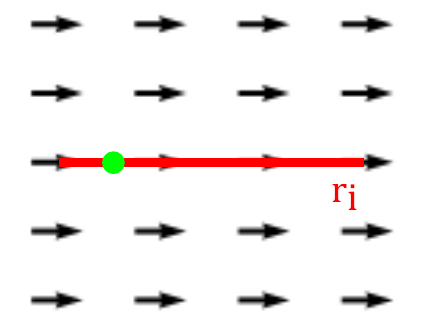}}
\hfil
\subfloat[]{\includegraphics[width=0.45\linewidth]{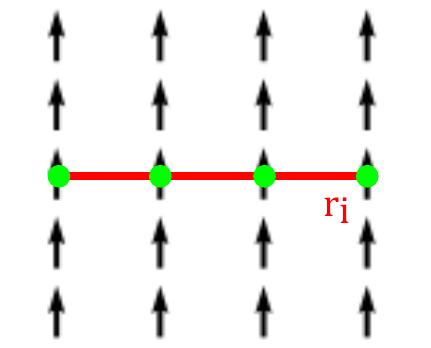}}
\caption{Schematic showing two extreme distributions of seed points (in green) on a curve $\bm{r}_i(s)$ (in red) of arc length $S = 3 l_s$ superimposed on the major eigenvector field $\bm{n}(s)$ (in black), in which: (a) the tangent to the curve and the major eigenvector are parallel, in which case $S'_i = 0$ and thus $\alpha = 1$, and (b) the tangent to the curve and the major eigenvector are orthogonal, in which case $S'_i = 3 l_s$ and $\alpha = 4 l_s$.}
\label{fig:seeding}
\end{figure}

Approximating an optimal distribution of seed points along the curves $\bm{r}_i(s)$ resulting from the topological template requires comparing the unit tangent vector to the curve, $\bm{t}(s)$, to the local alignment, represented by the major eigenvector $\bm{n}(s)$ of the alignment tensor field (along the curve).
For example, if $\bm{t}(s)$ is always parallel to $\bm{n}(s)$ then the curve $\bm{r}_{i}$ lies along a hyperstreamline and only a single seed point is needed anywhere within the curve, regardless of the desired spacing $l_s$ (Figure~\ref{fig:seeding}a).
The other extreme is if $\bm{t}(s)$ is always orthogonal to $\bm{n}(s)$, which would require $\alpha = S_{i}/l_s + 1$ seed points equally distributed along the curve, where $S_{i}$ is the arc length of the curve (Figure~\ref{fig:seeding}b).

In the presented method, for every curve $\bm{r}_{i}(s)$ in the topological template a weighting function $w_{i}(s)$ is defined,
\begin{equation}
w_{i}(s) = 1 - \left|\bm{t}(s)\cdot\bm{n}(s)\right|
\end{equation}
By integrating this weighting function along the curve, a renormalized arc length $S'_{i}$ of the curve can be found:
\begin{equation}
S'_i = \int_0^{S_{i}} w_i(s) \mathrm{d}s
\end{equation}
The number of seed points to be distributed along the curve is then given by $\alpha = S'_i/l_s + 1$.
Finally, these $\alpha$ seed points are distributed at specific points $s_{i,j=1\rightarrow\alpha}$ along the arclength of the curve governed by the $\alpha-1$ constraints of the form:
\begin{equation}
l_s = \int_{s_{i,j}}^{s_{i,j+1}} w_{i}(s)\mathrm{d}s
\end{equation}

This procedure is repeated for all curves $\bm{r}_i(s)$ in the topological template which results in a set of seed points $\mathcal{P}_{i}$.
Note that the unit tangent for each curve must first be computed as a function of arc length:
\begin{equation}
\bm{t}(s) = \left\|\frac{\mathrm{d}\bm{r}}{\mathrm{d}s}\right\|^{-1} \frac{\mathrm{d}\bm{r}}{\mathrm{d}s}
\end{equation}
which is accomplished by using spatial interpolation of the curve using cubic splines \cite{SciPy}.

\subsection{Hyperstreamline Computation}\label{sec:hyperstreamline}

Once the set of seed points $\mathcal{P} = \bigcup\mathcal{P}_{i}$ is determined, hyperstreamlines are computed and rendered at every seed point.
Directionality of the hyperstreamline computation is constrained when seed points lie on curves that enclose degeneracies/defects (identified from the topological template).
For these seed points hyperstreamlines are computed only in the direction pointing away from the vertex/degeneracy.
This approach avoids computation of hyperstreamlines in the vicinity of regions in the alignment tensor field where the major eigenvector becomes degenerate.

Hyperstreamline computation and rendering was performed using the Visualization Toolkit \cite{VTK} (version 5.10.1). 
The algorithm which this library implements is as follows:
\begin{enumerate}
    \item Given an alignment tensor field in the form of an unstructured grid, the Jacobi eigendecomposition algorithm is used to solve for the eigenvectors and eigenvalues at every grid-point.
    \item Integration of eqn. \ref{eqn:streamline} at every seed-point is then performed using a second-order Runge-Kutta method and spatial interpolation between grid-points.
    \item The size of the cross-section and its orientation along each hyperstreamline is scaled appropriately with respect to the size of the domain in order to improve visibility.
Additional scalar field data, such as biaxiality or major eigenvalue, can be represented through coloring of the hyperstreamline surface.
\end{enumerate}
In this work, the major eigenvalue field was used for hyperstreamline coloring.

\section{Results and Discussion}\label{sec:res_and_disc}

\begin{figure*}[!]
\centering
\subfloat[]{\includegraphics[width=0.3\linewidth]{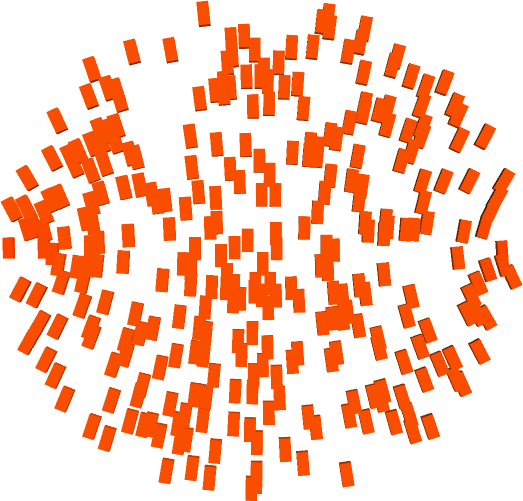}}%
\hfil
\subfloat[]{\includegraphics[width=0.3\linewidth]{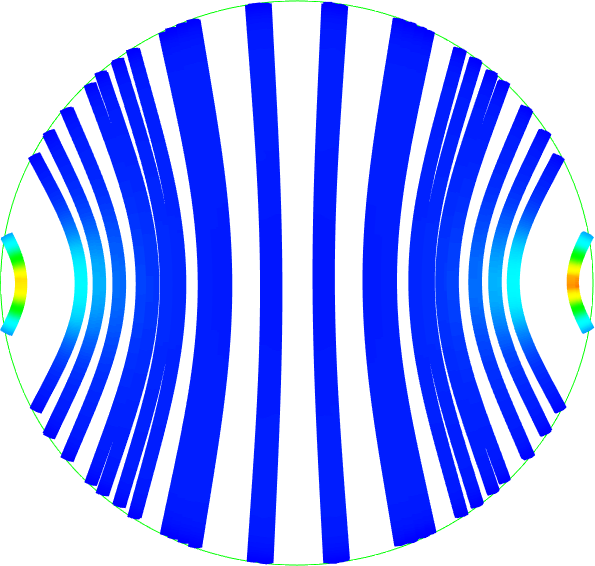}}%
\hfil
\subfloat[]{\includegraphics[width=0.3\linewidth]{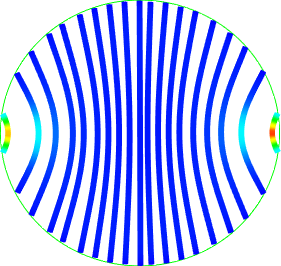}}%
\caption{Visualizations of a uniform defect-free alignment tensor field within a two-dimensional circular domain using: (a) rectangular tensor glyphs, (b) hyperstreamlines seeded uniformly in the field, (c) non-uniform seeding (Section \ref{sec:seed_distribution}) hyperstreamlines seeded along the boundary. The value of $l_{n}$ is $0.02$ in all cases and the domain diameter is $1$.}
\label{fig:circle_uniform}
\end{figure*}

The presented algorithm was applied to seeding three general types of two-dimensional alignment tensor fields observed in reorientation dynamics of nematic LCs: uniformly aligned defect-free domains, well-aligned domains with minimal defects present, and complex domains with many defects present.
These three cases represent the breadth of scenarios that might be encountered by a researcher.
In each case, domains are visualized using tensor glyphs and two different types of hyperstreamline seeding methods, in addition to the presented method.
The two alternative hyperstreamline seeding methods are used for comparison and are: uniform and boundary seeding.
In Sections \ref{sec:res:uniform} to \ref{sec:res:min_defects}, circular two-dimensional alignment tensor fields are used and in Section \ref{sec:res:many_defects}, a square domain is used.

In all three visualizations, the same value of the hyperstreamline spacing parameter $l_{s} = 2l_{n}$ is used, where $l_{n}$ is a characteristic length scale of the tensor field which governed by the physics of the problem (see Section \ref{sec:methods}).
Additionally, two optional parameters were included which were found to be useful for creating uncluttered hyperstreamline visualizations: the vertex seed radius and the vertex/edge seed ratio.
The vertex seed radius is the radius of the circle used as the curve $\bm{r}_i(s)$ around an orientational defect (vertex in the undirected graph) and was chosen to be $2.5l_n$ in the following visualizations.
The vertex/edge seed ratio specifies the relationship between the hyperstreamline spacing $l_{s}$ along boundaries/graph vertices versus along curves encompassing defects.
The value of this parameter used was $2$, which corresponds to the hyperstreamline spacing along edges (between defects) being twice that of the spacing around defects.
These two additional parameters are not data-specific, and thus the only input required from the user is the specification of $l_{s}$.

\subsection{Uniform Alignment}\label{sec:res:uniform}

Figures \ref{fig:circle_uniform}a-c show visualizations of an alignment tensor field within a circular domain using both rectangular tensor glyphs, uniform hyperstreamline seeding in the domain, and boundary hyperstreamline seeding using the seed distribution method described in Section \ref{sec:seed_distribution}.
In this case, the alignment tensor field contains no orientational defects and thus the presented method forms the topological template using the physical boundary (Figure \ref{fig:circle_uniform}c).

The tensor glyph visualization shown in Figure \ref{fig:circle_uniform}a uses a random distribution of points within the domain so that the scale of the glyph is large enough to be distinguishable.
In this simple case, using tensor glyphs results in a visualization that is indicative of the alignment tensor field configuration.

Using uniform hyperstreamline seeding, as shown in Figure \ref{fig:circle_uniform}c, results in a significant amount of hyperstreamline overlap in that multiple seed points are placed along lines of constant alignment.
Figure \ref{fig:seeding}a illustrates this scenario which results in a visualization that is severely cluttered.
Alternatively, the use of topology of the domain through seeding on its boundary (in the absence of defects) in combination with the presented seeding distribution method results in a hyperstreamline visualization (Figure \ref{fig:circle_uniform}c) that is approximately optimal, given the user-specified hyperstreamling spacing criteria.

\subsection{Minimal Defects in Alignment}\label{sec:res:min_defects}

\begin{figure*}[!]
\centering
\subfloat[]{\includegraphics[width=0.31\linewidth]{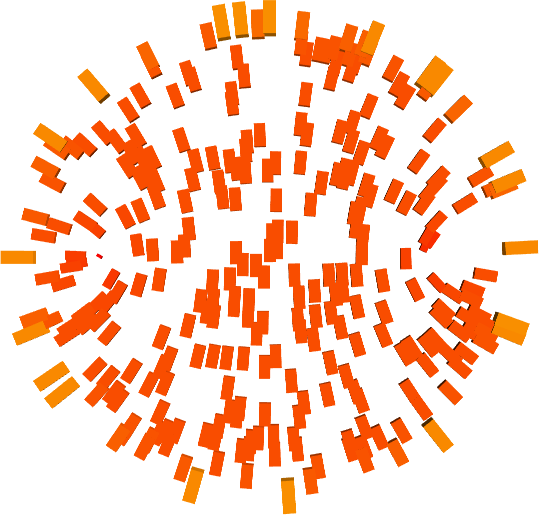}}%
\hfil
\subfloat[]{\includegraphics[width=0.31\linewidth]{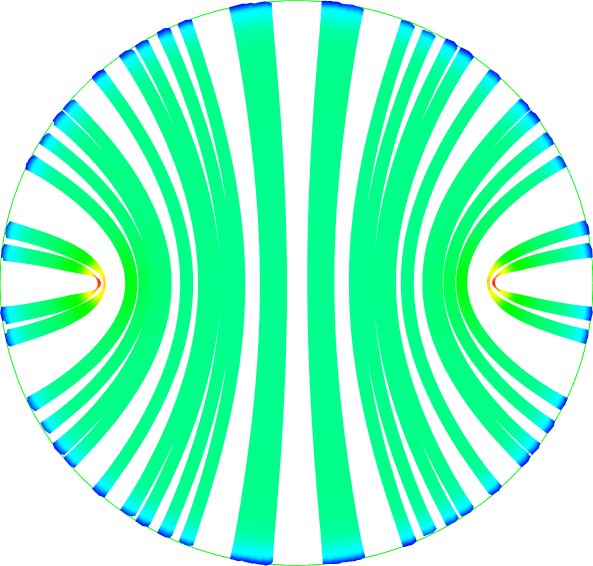}}%
\hfil
\subfloat[]{\includegraphics[width=0.31\linewidth]{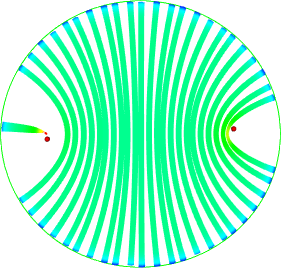}}
\hfil
\subfloat[]{\includegraphics[width=0.31\linewidth]{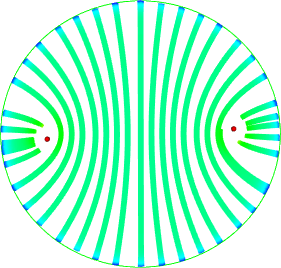}}%
\hfil
\caption{Visualizations of an alignment tensor field within a two-dimensional circular domain with minimal defects (shown in red) using: (a) rectangular tensor glyphs, (b) hyperstreamlines seeded uniformly in the field, (c) hyperstreamlines seeded along the boundary, (d) hyperstreamlines seeded along the undirected graph of defects (shown in Figure \ref{fig:topo_template}). The value of $l_{n}$ is $0.02$ in all cases and the domain diameter is $1$.}
\label{fig:circle_defects}
\end{figure*}

\begin{figure}[!]
\centering
\includegraphics[width=0.75\linewidth]{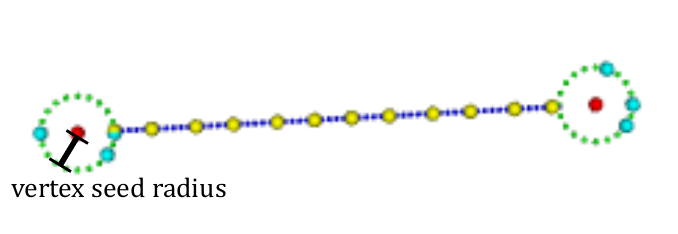}
\caption{Visualization of the undirected graph representing the topological template (Section \ref{sec:topo_template}) and seed points of the tensor field used in Section \ref{sec:res:min_defects}.  Seed points are shown in yellow and green, defects in red, and the curves comprising the template ($\bm{r}_i(s)$) are represented by dotted lines. The vertex seed radius is labelled in black. }
\label{fig:topo_template}
\end{figure}

Figures \ref{fig:circle_defects}a-d show visualizations of an alignment tensor field, now with two defects present, within a circular domain using both rectangular tensor glyphs and hyperstreamlines, respectively.
In this case, the presented method uses the undirected graph formed from the defects, shown in Figure \ref{fig:topo_template}, as opposed to the boundary.

Compared to the tensor glyph visualization (Figure \ref{fig:circle_defects}a), which again uses a random distribution of points within the domain, all three of the hyperstreamline visualizations provide a more understandable representation of the alignment tensor field.
In this case, the tensor field has significant gradients in alignment which is poorly represented in the glyph case.
This could be addressed by increasing the density of tensor glyphs proportional to the local spatial gradient in alignment, but this would result in an overlapping of glyphs and/or rescaling to the point that the glyphs are not distinguishable.
The advantage of using higher dimensional hyperstreamline visualization is clear in comparing Figure \ref{fig:circle_defects}a with Figures \ref{fig:circle_defects}b-d.
The continuous variation in direction that the hyperstreamline represents is not accessible with tensor glyphs.

Comparing Figures \ref{fig:circle_defects}b-d, using uniform seeding, boundary seeding with the presented seed distribution method, and the presented topological template/seed distribution method (Figure \ref{fig:topo_template}) demonstrates two significant disadvantages of both the uniform and boundary seeding methods.
First, both the uniform and boundary seeding methods admit the possibility of hyperstreamlines entering defect regions where degeneracies in alignment result in instabilities in the numerical method used to solve eqn. \ref{eqn:streamline}.
An example of this is shown in the left-center region of Figure \ref{fig:circle_defects}c where a hyperstreamline abruptly ends in the vicinity of a defect.
The hyperstreamline integration (Section \ref{sec:hyperstreamline}) failed in this region due to two eigenvalues having equivalent magnitudes.
The topologically-informed template results in seeding that avoids this failure-mode, as shown in Figure \ref{fig:circle_defects}c, in that computation of hyperstreamlines within the vertex/degeneracy seed radius is explicitly avoided.

The second disadvantage of using both the uniform and boundary seeding methods is that the spacing of hyperstreamlines is poorly constrained within the bulk of the domain.
In Figure \ref{fig:circle_defects}b, the same type of cluttering is observed as described in the previous case.
In Figure \ref{fig:circle_defects}c, the imposed spacing on the boundary is constrained well, but this results in cluttering of hyperstreamlines in the bulk of the domain.
The use of the topological template/seed distribution, shown in Figure \ref{fig:circle_defects}d, results in an approximately optimal hyperstreamline distribution such that the spacing is well-constrained in the bulk of the domain, while relaxing this spacing at the boundaries.
This hyperstreamline distribution is clearly preferred in that it results in an uncluttered visualization throughout the domain.

\subsection{Many Defects in Alignment}\label{sec:res:many_defects}

\begin{figure*}[!t]
\centering
\subfloat[]{\includegraphics[width=0.3\linewidth]{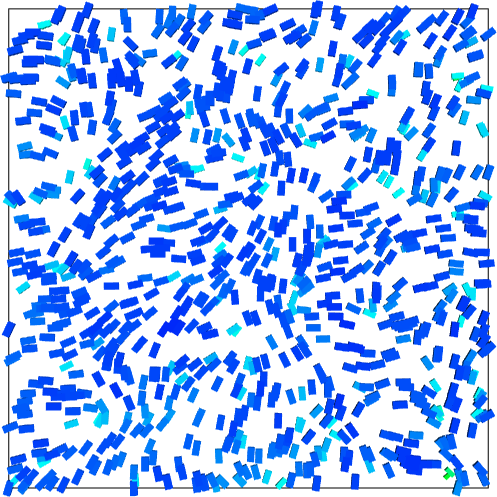}}%
\hfil
\subfloat[]{\includegraphics[width=0.3\linewidth]{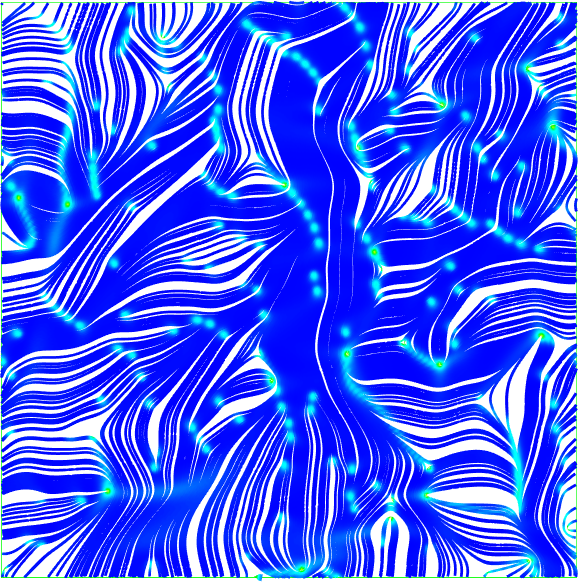}}%
\hfil
\subfloat[]{\includegraphics[width=0.3\linewidth]{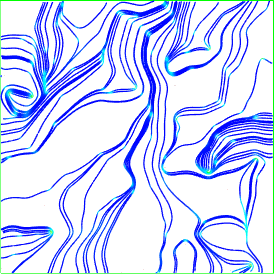}}
\quad
\subfloat[]{\includegraphics[width=0.3\linewidth]{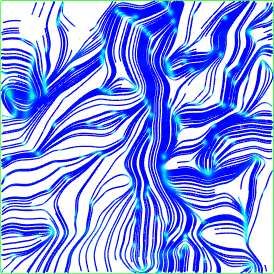}}%
\caption{Visualizations of a complex alignment tensor field within a two-dimensional square domain with many defects demonstrating: (a) the undirected graph of defects, (b) rectangular tensor glyphs, (c) hyperstreamlines seeded uniformly in the field, (d) hyperstreamlines seeded along the boundary, (e) hyperstreamlines seeded along the undirected graph of defects.
The value of $l_{n}$ is $0.008$ in all cases where the domain scale is $1$.}
\label{fig:square_defects}
\end{figure*}

Figure \ref{fig:square_defects}a-d show visualizations of a complex alignment tensor field with many defects.
This alignment tensor field would require analysis at multiple scales, including the largest scale (the whole domain) as is shown.
The significant disadvantage of tensor glyphs is apparent in comparing Figure \ref{fig:square_defects}a to Figures \ref{fig:square_defects}b-d; in order for the glyphs to be distinguishable, their scale must be large with respect to the characteristic length of variation in alignment.
Thus many important features of the alignment tensor field are not visualized due to the coarseness of the visualization.
Once again, the higher dimensional character of hyperstreamlines provides a more useful visualization in all cases.

\begin{figure}[!t]
\centering
\includegraphics[width=0.9\linewidth]{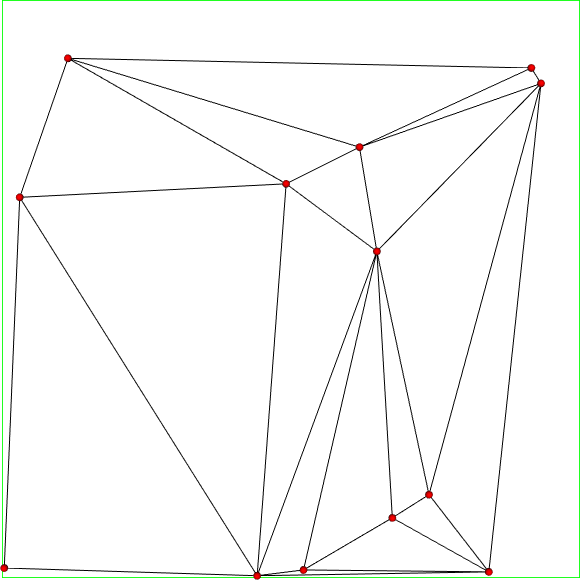}
\caption{Visualization of the undirected graph representing the topological template of the tensor field used in Section \ref{sec:res:many_defects}.}
\label{fig:topo2}
\end{figure}

Focusing on uniform hyperstreamline seeding, Figure \ref{fig:square_defects}b, the degree of hyperstreamline overlap is severe.
While visualization of the largest scale of the alignment tensor field is distinguishable, analysis at smaller scales is infeasible due to overlap.
Comparing boundary and topological template seeding, Figure \ref{fig:square_defects}c-d, it is observed that both disadvantages described in the previous section are magnified for boundary seeding in this larger and more complex alignment tensor field.
Through the use of the topological template, shown in Figure \ref{fig:topo2}, these features are accounted for and the resulting visualization is meaningful on multiple scales, ranging from the whole field to sub-regions.

\section{Conclusion}\label{sec:conclusions}

A topologically-informed method is presented for seeding of hyperstreamlines for visualization of alignment tensor fields.
The method is shown to approximate an optimal distribution of hyperstreamlines for a breadth of two-dimensional alignment tensor fields ranging from those without defects to those with complex topology.
The method requires only a single parameter from the user, avoids possible failure modes in hyperstreamline computation, and requires no iteration to yield satisfactory hyperstreamline spacing.
The results of applying the presented seeding method show that it enables automated efficient hyperstreamline-based visualization of alignment tensor fields and thus enhances the ability of researchers to interpret this type of data.
While the description of the method and results were limited to two-dimensional tensor fields, the presented work provides a basis for extension of the seeding method to three-dimensional alignment tensor fields.

\section*{Acknowledgment}

This research was supported by the Natural Sciences and Engineering Research Council (NSERC) of Canada and the facilities of the Shared Hierarchical Academic Research Computing Network (SHARCNET).

\bibliographystyle{IEEEtran}
\bibliography{self_assembly,computational,general}

\end{document}